\documentclass [12pt]{article}
\usepackage{graphicx}

\title{Acoustics of the Indian Drum}
\author{ Siddharth S Malu and Advaith Siddharthan}

\begin{document}

\maketitle

\tableofcontents

\begin{abstract}
Prof.  Raman found that the Indian drums all
produced harmonics, whereas the western drums do not, a fact
that is proved theoretically.  He also found that that the
density distribution present as a central black patch on the
Indian drums is {\em not} of the type $r^{-n}$.  However,
R.Siddharthan, P.Chatterjee and Vikram Tripathi have stated
in their paper \footnote{see {\em Physics Education} Oct. 1994.} that they found the harmonics of the Indian
Drums all in order, except the fundamental, which was higher
than what it should have been.  This article presents two
theoretical models (or distributions) one of which was found to yield
frequency ratios identical to those found by R.Siddharthan
et al.
\end{abstract}

\begin{figure}\label{fig:tablaset}
\begin{center}
\includegraphics*[height=14cm,width=17cm,angle=0]{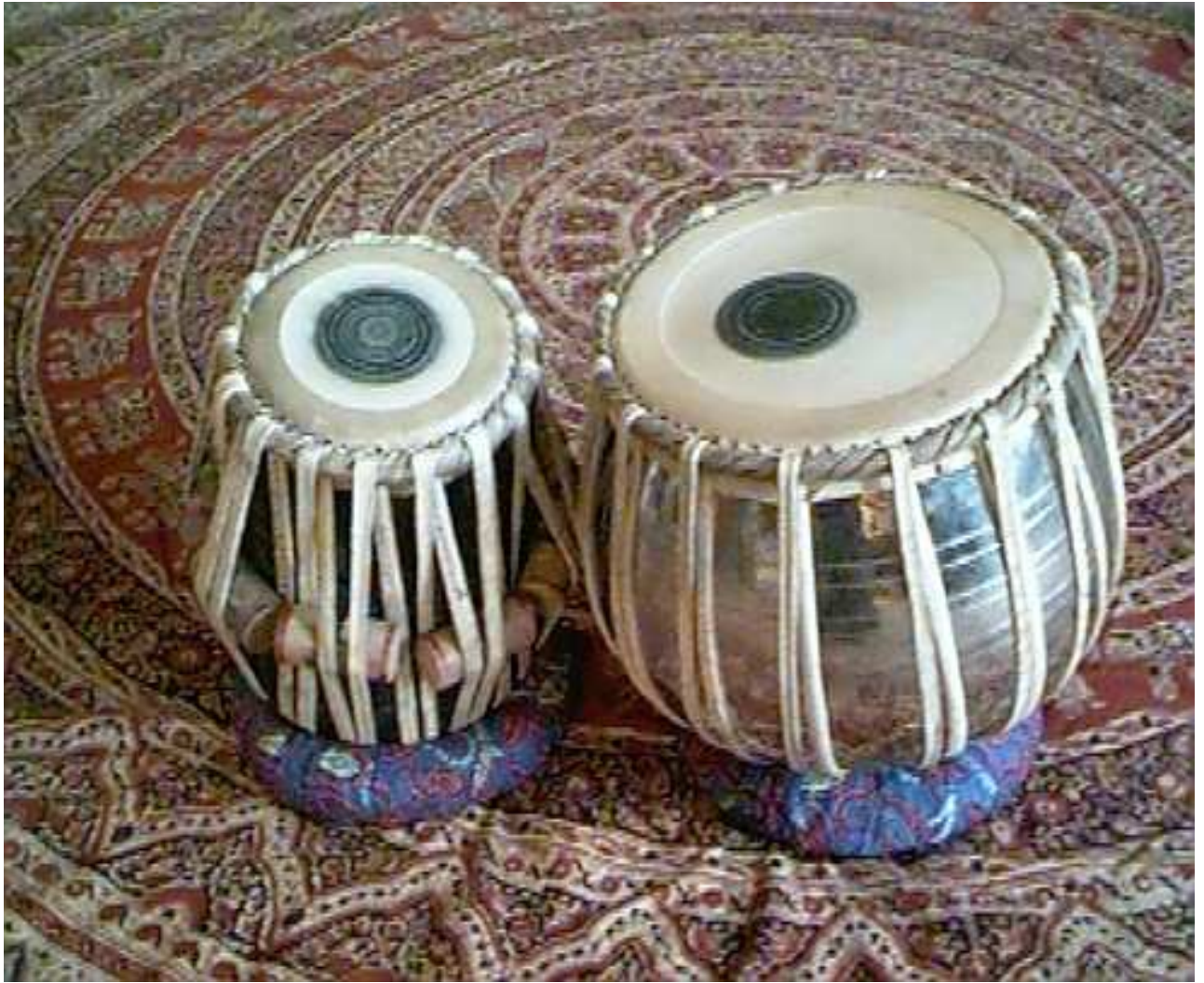}\\
\end{center}

\caption{A view of the Tabla (photo courtesy of TABLA.COM, Copyright
1999). This article is concerned with the right-hand side tabla, which
can be seen in this figure on the left.}
\end{figure}

\section{Introduction - Music, Harmonics and Overtones}
Sound is caused by alternate compressions and rarefactions of pressure in the air. The ear-drum vibrates with these so-called longitudinal waves, and that is how the sense of sound is had. Any arbitrary sound, however, does not please the ear. The normal human ear can, in general, differentiate between what is pleasing and what is noise. This section gives the important differences between Music and Noise.
 
\subsection{Music}
Music has its own definite wave patterns, so that any arbitrary sound cannot be called music. One can, by looking at the wave patterns, find out whether a given sound is musical or not. Music has the following two attributes that distinguish it from noise:
\begin{enumerate}
\item {\bf Pitch:} The pitch of a wave pattern characterises the time after which the wave pattern repeats itself. For the sensation of pitch, therefore,{\em {\bf it is essential that the waveform be periodic}}. A non-periodic waveform will not give the sensation of pitch, and can hence be labeled {\em noise}.

\item {\bf Timbre:} Timbre refers to the quality of sound. A tuning fork, a {\em sitar}, a {\em tanpura} and a piano may all have the same pitch, yet their timbre may be completely different. {\bf Pitch} refers to the {\bf frequency} of repetition of the waveform, whereas {\bf timbre} depends on the {\bf shape} of the waveform. What timbre really refers to is the relative amplitudes of the various component {\em single} frequencies that make up a particular sound wave. 

\end{enumerate}   

\subsection{Overtones and Harmonics}
Any complex waveform may be expressed as a linear superposition of a number of sine waves of different frequencies. Notes produced by musical instruments consist of sine waves of one fundamental frequency and of higher frequencies called {\bf overtones}. For the resultant waveform to be periodic, the overtones have to be integral multiples of the fundamental frequency. Overtones that are integral multiples of the fundamental frequency are called {\bf harmonics}. 

Unless a majority of overtones in a particular sound are harmonic, the
waveform will not be periodic and will not have a discernible
pitch. {\em{\bf Thus for a note to sound {\em musical} \footnote{{\em Musical},
as used here just means having a definite pitch, which is very
important in Indian Classical accompaniments; however, this does not
mean that the Western drums are 'non-musical', they produce rhythm,
just like their Indian counterparts, see section 1.3.}, a majority of the overtones must be harmonic}}.  

\subsection{Indian and Western Drums}

The Tabla is the most well-known of all Indian drums. Amir Khusro created it when he split the ancient Indian drum, the Pakhawaj into two parts. Of these, the right drum has a black patch in the centre, as mentioned earlier. This is made of a mixture of iron, iron oxides, resin, gum etc. and is stuck firmly on to the membrane. The thickness of this patch decreases radially outwards.

The left hand drum has a wider membrane, and has a black patch similar to the one in the right drum, except that it is unsymmetrically placed on one side of the membrane. It is worth noting here that the left hand drum is not used to produce harmonics but to provide lower frequencies in the overall sound while the Tabla is being played.

The key difference between Indian and western drums is the absence of the central loading in the case of the Western drums. It is shown mathematically in section 4 that a uniform circular membrane {\em cannot} produce harmonics. In section 5, we then investigate how the drum may be made harmonic by considering two theoretical models, i.e. radial density distributions.

{\bf Thus, this is a study of how the density variation of the
membrane affects the frequencies of the overtones.} 

Indian Classical music is such that a recital consists (usually) of a
single performer, and a couple of accompaniments. In this situation,
the accompaniment (usually a drum, the {\em Tabla} for instance, {\em
must} be harmonic, since otherwise, the aharmonicity would completely
disrupt the recital. 

Western drums, however, play a very different part in Western
Classical music. They provide a rhythm to the music produced by the
rest of the instruments, and (strictly speaking) do not need to produce
harmonics. 

\section{The Wave Equation and its solution} 
The wave equation, in the case of a membrane, describes the relationship between the displacement of a point on the membrane changes with position and with time, when it is disturbed in some way.  

The general form of the Wave equation is :
\begin{equation} 
\nabla^{2}u = \frac{1}{c^{2}}
\frac{\partial^{2}u}{\partial^{2}t}.
\end{equation}

For a circular membrane, in polar co-ordinates, the wave equation may be written as
\begin{equation}
\frac{\partial^{2} u}{\partial r^{2}} + \frac{1}{r} \frac{\partial u}{\partial r} + \frac{1}{r^{2}} \frac{\partial^{2} u}{\partial \theta^{2}} = \frac{1}{c^{2}} \frac{\partial^{2}u}{\partial t^{2}}
\end{equation}
where $(r,\theta)$ describes any arbitrary point on the membrane that is assumed to have:
\begin{enumerate}
\item Uniform mass per unit area
\item Uniform tension per unit length
\end{enumerate}
and $u(r,\theta,t)$ is the vertical displacement of the point at a time t. 
Also,
\begin{equation}
c^{2} = \frac{\tau}{\rho}
\end{equation}
where c is the speed at which the wave travels on the membrane, $\tau$ is the tension per unit length in the membrane and $\sigma$ is the mass per unit area of the membrane.

This equation is solved using the technique of separation of variables, where it is assumed that 
\begin{equation}
u(r,\theta,\phi) = R(r) \Theta(\theta) T(t).
\end{equation}
Here, $R$ depends only on $r$, $\Theta$ only on $\theta$, and $T$ only on $t$. Making this substitution, and solving, the three quantities $T$, $\Theta$ and  $R$ are found to be:
\begin{enumerate}
\item {\bf The $T$ solution}: This is a simple sine (or cosine) function, indicating that the membrane undergoes simple harmonic oscillations in time. Its general form is:
\begin{equation}
T = cos\left(\omega t + \phi \right),
\end{equation}
where $omega = ck$ and $\phi$ is a phase which depends on the initial conditions. However, it can be set to zero without loss of generality.  

\item The {\bf $\Theta$ solution}: This tells us how $u$ changes with angle, i.e. how displacement changes as we move along a circular path of fixed radius. This part is:
\begin{equation}
\Theta = cos(m \theta+\psi).
\end{equation}
$\psi$ is again a phase which depends on initial conditions, and may be put to zero without loss of generality. There are certain values of $\theta$ for which $\Theta(\theta)$ reduces to zero. Thus, $u$ would, at all times be equal to zero on all diameters along which $m\theta$ is an integral multiple of $\frac{\pi}{2}$. 

{\bf {\em Nodal Diameters}} of a drum are those diameters that remain stationary while the rest of the drum is vibrating. For a particular value of $m$, there are $m$ symmetrically placed Nodal Diameters.

\item {\bf The $R$ solution}: This describes how the displacement of the membrane above the plane of rest (i.e. $u$) changes as we move outward, along a radius. There are certain circles that remain stationary on the membrane while the rest of the drum vibrates. This happens whenever the function $R(r)$ crosses the value zero. These circles are called {\bf {\em Nodal Circles}} and their radii correspond to those values of $r$ at which $R(r)$ reduces to zero. The $R$ solution is given by:
\begin{equation}
R(r) = J_{m}(kr),
\end{equation}
where $J_{m}$ is the Bessel function of the first kind of order $m$. 
\end{enumerate}
A few Bessel functions are plotted in Fig.2 and Fig.3.

\begin{center}
\begin{figure}\label{fig:Bessel1}
\includegraphics*[height=10cm,width=10cm,angle=-90]{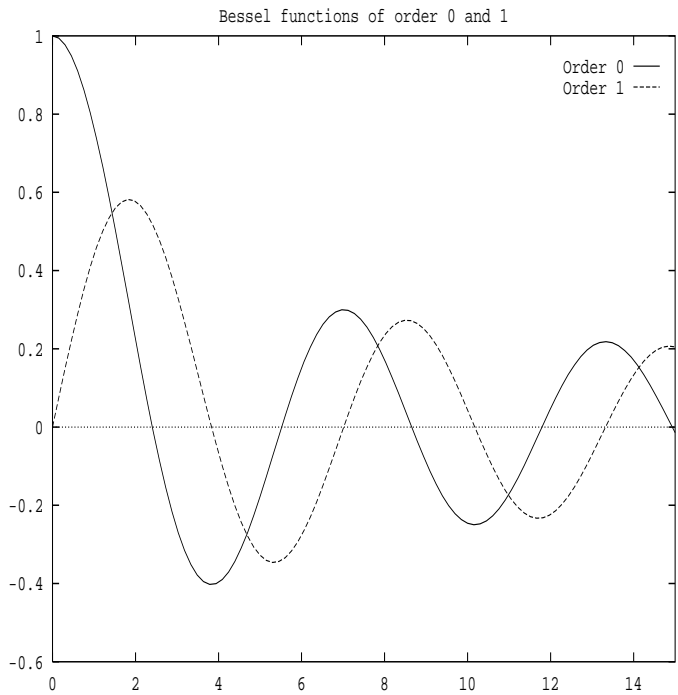}\\

\caption{Bessel Functions of order 0 and 1.}
\end{figure}
\end{center}

\begin{center}
\begin{figure}\label{fig:Bessel2}
\includegraphics*[height=10cm,width=10cm,angle=-90]{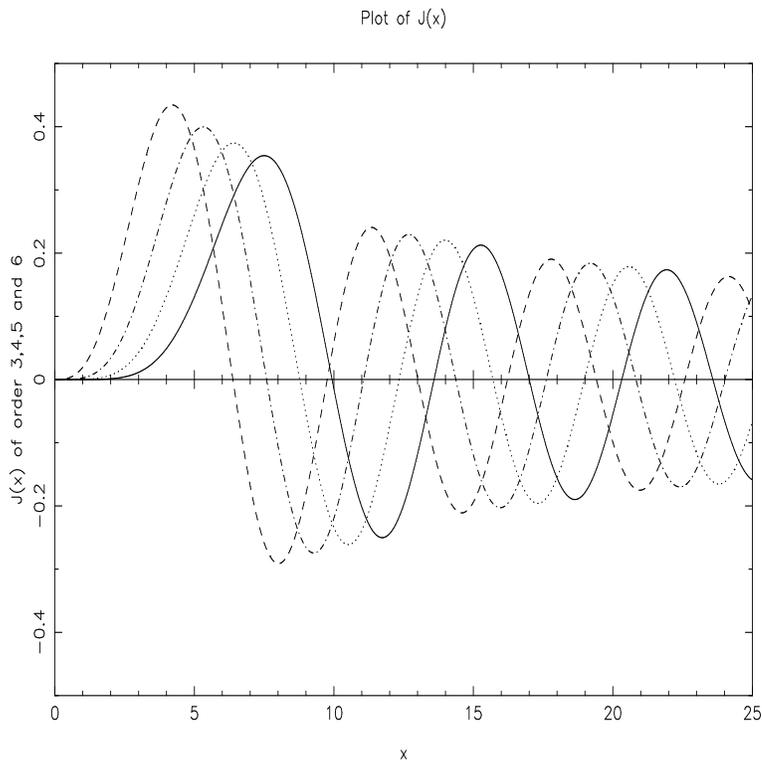}\\

\caption{Bessel Functions of order 3,4,5 and 6.}
\end{figure}
\end{center}

\subsection{Normal Modes}
The complete solution may be written as:
\begin{equation}
u(r, \theta, t) = \sum J_{m}(k_{mn}r) cos(m\theta) cos (\omega_{mn}t)
\end{equation}

The most general solution is thus a superposition of various different modes, where in each mode, the whole drum vibrates with one frequency, has $m$ Nodal Diameters and $n-1$ nodal circles. These modes are called the {\bf {\em Normal Modes}} of the drum. The first nine Normal Modes are shown in Fig. 4.
\begin{center}
\begin{figure}\label{fig:Normalmodes}
\includegraphics*[height=14cm,width=13cm,angle=-90]{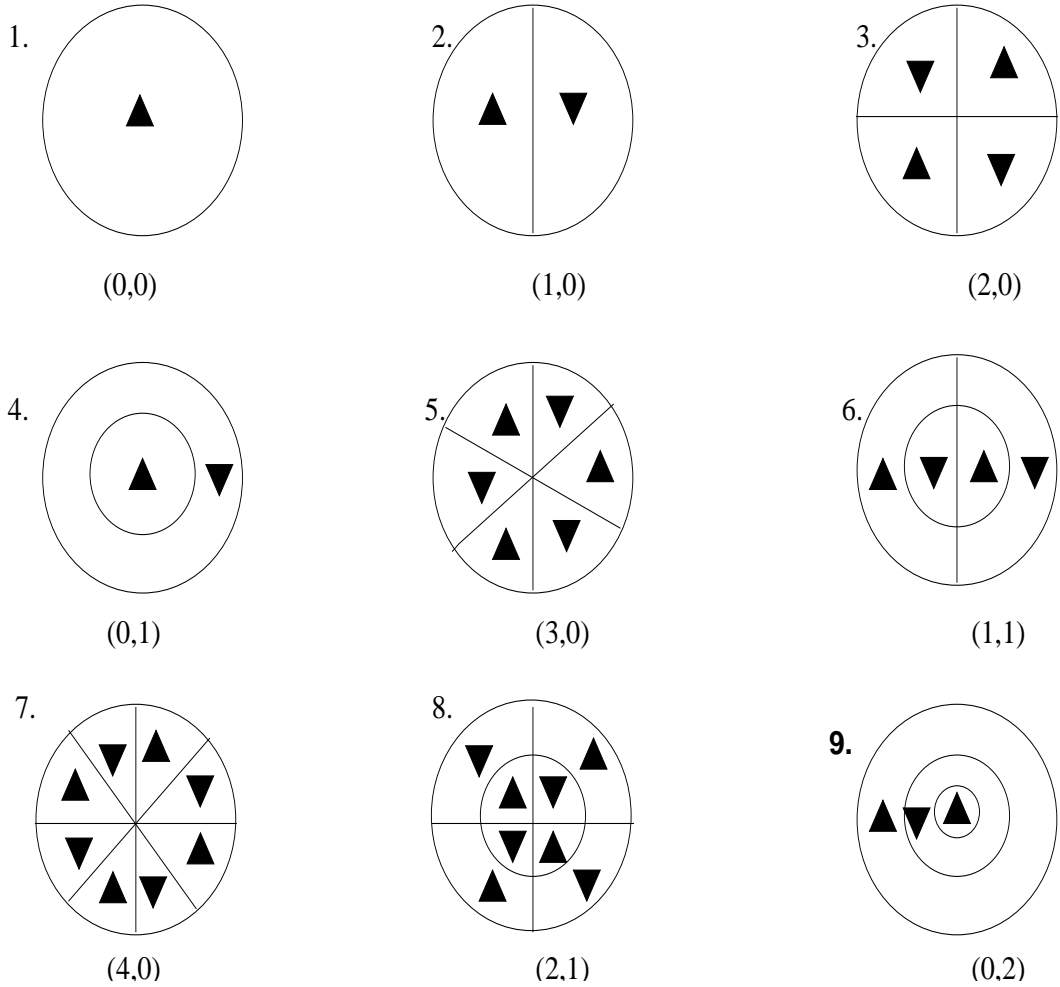}\\

\caption{The first nine Normal Modes of a circular membrane. Quantities in brackets are ({\em Nodal Diameters}, {\em Nodal Circles}). Arrows indicate the direction in which the various parts, separated by Nodal circles and diameters, vibrate.} 
\end{figure}
\end{center}

\subsection{The Boundary condition}
The membrane is bound in the form a circle over the head of the wooden shell. Therefore, the periphery is always stationary. This has to be taken into account when the solution is being calculated. The condition may be stated mathematically as follows:
\begin{equation}
u(r,\theta,t) = 0 
\end{equation}
at $r = a$
or
\begin{equation}
J_{n}(k_{mn}a) = 0.
\end{equation}
That is, the boundary of the membrane should correspond to one of the zeroes of the Bessel function of any order $n$. This will yield the values of frequencies that are 'allowed' by the boundary condition. However, we are interested only in the relative ratios of these frequencies. 

Let us quantify the 'allowed' frequencies. Let $b_{mn}$ be the mth zero of the Bessel function of order $n$. Then the above boundary condition becomes:
\begin{equation}
k_{mn} = \frac{b_{mn}}{a},
\end{equation}
i.e. the allowed frequencies are proportional to the zeroes of Bessel functions. As noted earlier, however, the values of these zeroes have no integral ratio to each other, so that the ratios of the frequencies are also non-integral.

Thus, {\bf {\em the simple circular membrane, and hence also Western drums, cannot produce harmonics.}}

\section{Making the Drum Harmonic}
Indian drums, in general, play a very different role as accompaniments
in Classical music as compared to their western counterparts, as
mentioned in sections 1.2 and 1.3. Since a uniform membrane does not give us harmonics, we tried solutions to a membrane with a density variation. The simplest possibility is a loading which varies only with $r$. We looked at the right hand tabla and tried various symmetric density distributions that resembled the actual loading. We describe  here the one that was found to be successful. But first, we note the change produced in the wave equation.

Recall that $c^{2} = \frac{\tau}{\rho}$. Now, however, $\rho = \rho(r)$, so that  $c^{2}(r) = \frac{\tau}{\rho(r)}$.
This causes the wave equation to change from 
\begin{equation}
\frac{d^{2}R}{dr^{2}} + \frac{1}{r}\frac{dR}{dr} + \left(\frac{\omega^{2}}{c^{2}} - \frac{m^{2}}{r^{2}}\right)R = 0 
\end{equation}
for a uniform membrane to 
\begin{equation}
\frac{d^{2}R}{dr^{2}} + \frac{1}{r}\frac{dR}{dr} + \left(\frac{\omega^{2}}{c^{2}(r)} - \frac{m^{2}}{r^{2}}\right)R = 0 
\end{equation}
or, equivalently,
\begin{equation}
\frac{d^{2}R}{dr^{2}} + \frac{1}{r}\frac{dR}{dr} + \left(\frac{\rho(r)\omega^{2}}{\tau} - \frac{m^{2}}{r^{2}}\right)R = 0 
\end{equation}
for a loaded membrane.

\begin{center}
\begin{figure}[h]\label{fig:loading}
\includegraphics*[height=14cm,width=12cm,angle=-90]{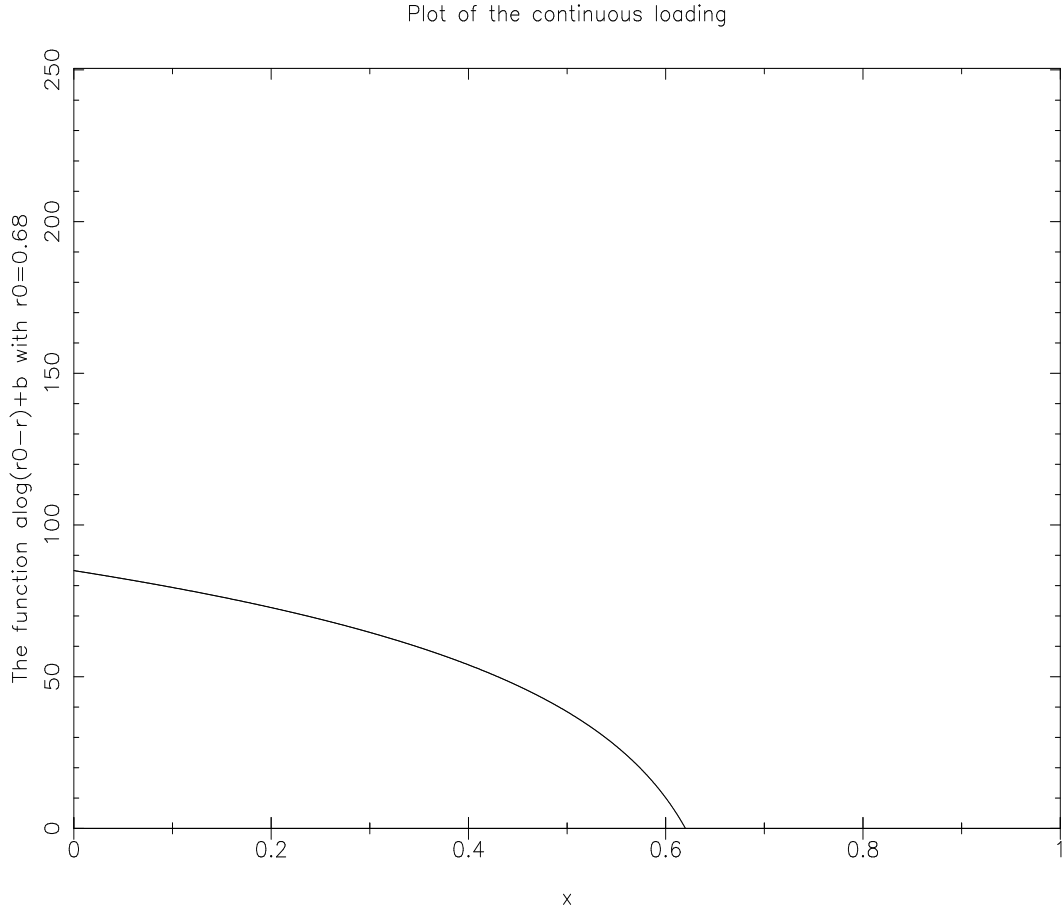}\\
\caption{A side view of the loading, as is seen on the Tabla, represented as the function $a log(r_{0}-r)+b$.} 
\end{figure}
\end{center}

\begin{center}
\begin{figure}[h]\label{fig:loading2}
\includegraphics*[height=14cm,width=12cm,angle=-90]{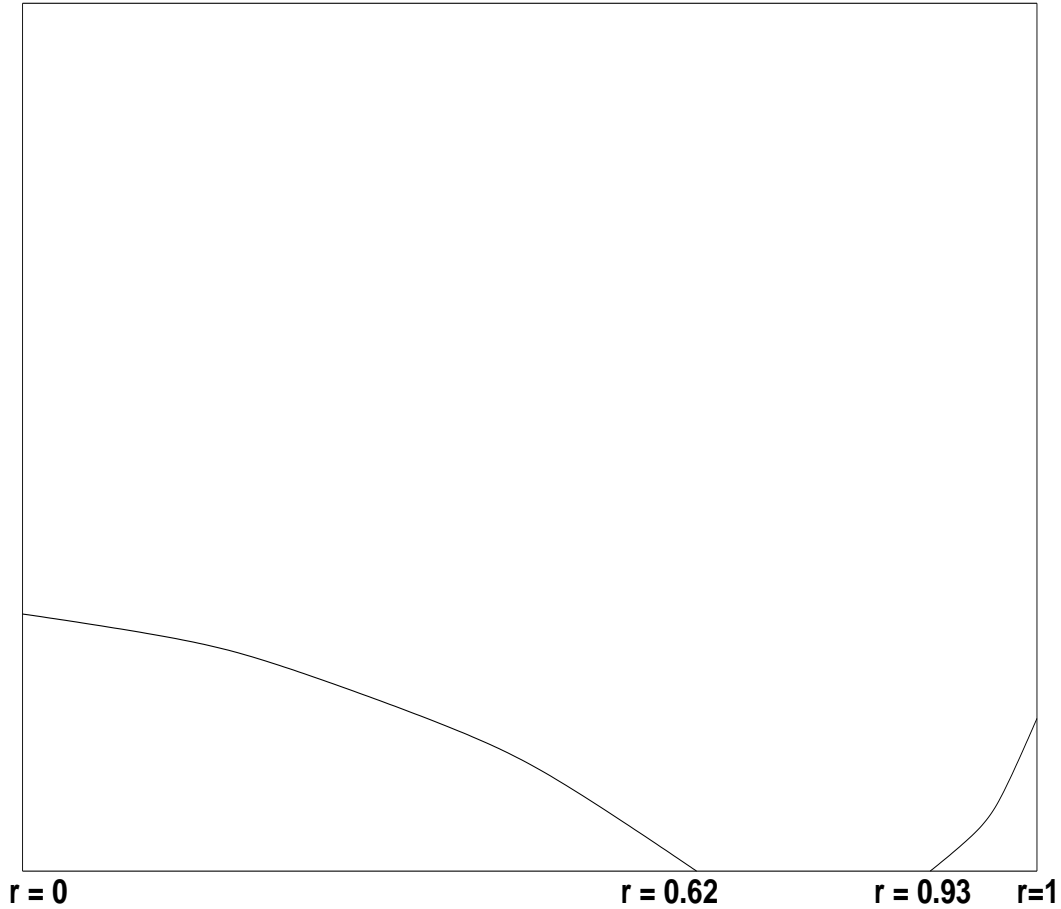}\\
\caption{The actual function used.} 
\end{figure}
\end{center}

The Radial equation can be solved for various distributions. We tried two kinds of loading patterns:
\begin{enumerate}
\item {\bf The Step function}: Concentric rings with varying density
were considered. This gave good results, as can be seen in the table.
\item {\bf The Continuous Loading}: The actual loading on the tabla is a step function to begin with. However, even though it is stuck in parts, the loading becomes more or less continuous after the tabla has been played for some time. The loading therefore looks quite like the function plotted in Fig. 5. 
\end{enumerate}
It was also found that if an exponential function is included toward
the periphery, the results obtained for both the distributions
improve. The form of the function used, along with the continuous loading, was $c e^{d r}$, where $c,d$ are constants.
The actual function used as loading is shown in Fig. 6.

\section{Calculation and results}
Both the distributions given in the previous section were put into the Radial part of the wave equation, and the equation was solved numerically using the second-order Runge-Kutta method.

In writing down the wave equation for the loaded membrane, the following two assumptions were involved:
\begin{enumerate}
\item Normal Modes exist even in the loaded membrane
\item Tension per unit length is the same throughout the membrane, even in the loaded part.
\end{enumerate}

\subsection{Calculation}
It may be seen from figure that the function chosen for Continuous loading varies slowly with $r$ near the centre of the membrane. Also, in the discrete loading, the density remains constant in a particular concentric circle, so that that the density is not varying in the centre. Due to this reason, the initial conditions (for starting the solution of the wave equation by the Runge-Kutta method) were assumed to be identical to those of the Bessel functions. 

The allowed frequencies were found in the following way:
First, the Radial equation was written as 
\begin{equation}
\frac{d^{2}R}{dr^{2}} + \frac{1}{r}\frac{dR}{dr} + \left(\rho(r)k'^{2} - \frac{m^{2}}{r^{2}}\right)R = 0 
\end{equation}
where
\begin{equation}
 k'^{2} = \frac{\omega^{2}}{\tau}.
\end{equation}  
This is the equivalent of the Bessel equation of order $m$ for the loaded case.

Now, keeping $m=0$, the value of $k'$ was varied and the solution
plotted as a graph on the screen until its value became zero at the
boundary. The corresponding value of $k'$ was noted. $k'$ was then
increased, until the solution again became zero, but this time the
solution passed through zero once, meaning that there was a nodal
circle. The process was continued for a particular value of $m$. Then,
$m$ was increased by one, and the same process was continued, and all
the allowed frequency values were noted.

\subsection{Problems with the calculation}
The method mentioned in the previous section works fine for $m=0$ and $m=1$, since the initial values of the Bessel functions and their derivatives are non-zero for the zeroth and first orders. However, for $m=2$ and higher orders, values of both the Bessel function and its derivative become zero at $r=0$. This leads to the solution becoming zero at all points for $m=2$ and higher orders.

This happens because the Runge-Kutta method depends on the initial value of the solution and its derivative (i.e. the value at $r=0$). Since the iteration begins with the derivative and the initial value as zero, it continues to be so for further values of $r$. 

To get around this problem, iteration was begun not from $r=0$ but from an infinitesimally small value, in this case $r=0.0001$. The values of the Bessel function and its derivative at this point were computed from the series expansion of the Bessel function, considering the first four terms. The rest of the procedure was as in the previous section, but this 'initial value' had to be calculated for each order separately. 

\subsection{Results}
The most surprising result we got was that for the continuous loading,
all the overtones were found to be nearly harmonic, but the
``fundamental'' itself was higher than what it should have been. In
particular, the ratios were found to be 1.07:2:2:3 etc. These results
are summarized in Table 1.The base in column 3 is the 2nd normal mode,
with 1 nodal diameter. This is done to illustrate the fact that the
fundamental is absent and the lowest eigenvalue is 1.07 times the fundamental.

\begin{table}[h]
\begin{center}
\begin{tabular}{|c|c|c|c|c|} \hline\hline
{\em Normal Mode} \footnote {see Fig. 4} & \multicolumn{4}{c|}{\em Frequency ratio} \\ \hline
(Nodal diameters, Circles) & Unloaded & Continuous Loading & Multiple rings & Tabla \\ \hline
1(0,0) & 1.00 & 1.07 & 1.00 & 1.00 \\ \hline
2(1,0) & 1.59 & 2.00 & 1.96 & 2 \\ \hline
3(2,0) & 3.14 & 2.98 & 2.98 & 3 \\ \hline
4(0,1) & 2.30 & 2.99 & 3.03 & 3 \\ \hline
5(3,0) & 3.65 & 4.00 & 4.02 & 4 \\ \hline
6(1,1) & 2.92 & 4.00 & 3.95 & 4 \\ \hline
7(4,0) & 3.16 & 5.01 & 5.02 & 5 \\ \hline
8(2,1) & 3.50 & 5.01 & 5.00 & 5 \\ \hline
9(0,2) & 3.60 & 5.02 & 4.80 & 5 \\ \hline
10(1,2) & 4.24 & 6.02 & 5.20 & - \\ \hline
11(1,3) & 5.55 & 7.80 & 7.03 & - \\ \hline
12(2,2) & 4.85 & 7.00 & 5.90 & - \\ \hline
13(3,1) & 4.06 & 6.04 & 6.02 & - \\ \hline
14(4,1) & 4.60 & 7.09 & 7.05 & - \\ \hline
\end{tabular}
\end{center}
\end{table}

Thus, a theoretical model with the first seven harmonics is possible, which also resembles the actual loading pattern on the Indian drums. However, with the kind of loading that the Indian drums have, the fundamental is absent and a slightly higher note is present instead. This is in accordance with recent observations of the frequencies present in the Indian drums.

\section{Discussion} 
The results table shows certain modes with relative frequency ratios that differ from the exact harmonic ratios by small amounts (e.g. 0.01, 0.02). However, the minimum difference required between two frequencies so that a normal human ear can distinguish between them is $6-7 Hz$. Considering that the fundamental note in Indian Classical ({\em Sa} of the base {\em Saptak}) is $240 Hz$, these differences work out to $2.4 Hz$ to $5 Hz$. Thus, the human ear is not able to make out these differences. 

\pagebreak

{\bf {\em Acknowledgements}}\\
This investigation was done as a project at St. Stephen's College,
Delhi, and was sponsored by the Deptt. of Science and technology,
Govt. of India, and was exhibited, along with other projects done at
St. Stephen's College, at the 85th session of the Indian Science Congress at Hyderabad, 3rd to 10th January, 1998.

We would like to thank Dr. S.C. Bhargava, Mr. N. Raghunathan,
Dr. B. Phookun, Dr. S. Aggarwal and Mr. S. Grewal of St. Stephen's
College for valuable suggestions and guidance during the project
period.


\begin{thebibliography}{99}
\bibitem{ad1} Tripathi, V., Siddharthan, R. and Chatterji, P. {\it Physics Education}, Oct. 1994.
\bibitem{ad2} ed. Ramaseshan, S. {\it Collected Papers of C.V. Raman},
1988. Bangalore: Indian Academy of Sciences.
\bibitem{ad3} Arfken, G. {\it Mathematical Methods of Physics}.
\end{thebibliography}
\end{document}